# #Coronavirus or #Chinesevirus?!: Understanding the negative sentiment reflected in Tweets with racist hashtags across the development of COVID-19

Xin Pei[1]  Deval Mehta[2]

[1]Nanyang Technological University   [2]IBM Research Australia

## Abstract

Situated in the global outbreak of COVID-19, our study enriches the discussion concerning the emergent racism and xenophobia on social media. With big data extracted from Twitter, we focus on the analysis of negative sentiment reflected in tweets marked with racist hashtags, as racism and xenophobia are more likely to be delivered via the negative sentiment. Especially, we propose a stage-based approach to capture how the negative sentiment changes along with the three development stages of COVID-19, under which it transformed from a domestic epidemic into an international public health emergency and later, into the global pandemic. At each stage, sentiment analysis enables us to recognize the negative sentiment from tweets with racist hashtags, and keyword extraction allows for the discovery of themes in the expression of negative sentiment by these tweets. Under this public health crisis of human beings, this stage-based approach enables us to provide policy suggestions for the enactment of stage-specific intervention strategies to combat racism and xenophobia on social media in a more effective way.

## Introduction

On 16th March 2020, a post from the official Twitter account of Donald Trump, the current president of the United States of America, stirred up a tremendous wave of debates across the globe. In this tweet, he referred to Coronavirus as Chinese virus. Ironically, this label that embodies a sense of racism and xenophobia becomes a new popular hashtag - #Chinesevirus, appearing in numerous posts involved in public discussions over COVID-19 on Twitter. The popularity of #Chinesevirus is accompanied by the proliferation of many other racist hashtags that either reflect the conflation of coronavirus with racial cultural identities, such as #Kungflu, or aim at delivering the message of racial exclusion, such as #sinophobia. This worldwide emergent phenomenon is calling for scholarly attention to the rising racism and xenophobia expressed in the public spheres of social media.

This brings us to a classic question - while social media have been acting as a significant equalizer that endows people with the fundamental right of free speech regardless of their class, gender, nationality, and race, whether the very same platforms are becoming fields where - in the garb of free speech – bias and hatred are able to flourish (Jakubowicz 2017; Lara-Cabrera et al. 2017; Thompson 2011)? More than once, history has shown us that when it comes to race issues, social media could act as the breeding ground of the expression of racism and xenophobia where rationale information sharing might be marginalized (Alkiviadou 2019; Ben-David and Matamoros-Fernández 2016; Chinnasamy and Manaf 2018). In the case of 2017 Jakarta Gubernatorial Election, supporters of competing camps with leaders from different races deployed social media as the battleground for mutual attacks, subsequently transforming the freedom of speech into the freedom to express hate (Lim 2017). In India, every day, countless statements of racial discrimination against Muslims appear on different social media platforms (Mirchandani 2018). Under some circumstances, the social media mediated diffusion of racism and xenophobia has been unfortunately evolved into physical violence and regional conflicts in the off-line world (Hubi 2019).

It is within this context that our attention is drawn to the analysis of tweets marked with racist hashtags under the topic of COVID-19. Especially, we focus on the tweets that reflect the negative sentiment. This is due to the reason that racism and xenophobia are more likely to be expressed negatively (Redlawsk, Tolbert, and McNeely 2014; Tulkens et al. 2016). Not limited to recognizing the negative tweets with racist hashtags, this study develops a stage-based sentiment analysis to map the changing negative sentiment expressed in these tweets across different development stages of COVID-19, where this disease evolved from a domestic epidemic to a public health emergency of international concerns and later to the global pan-demic. Additionally, keywords of the negative tweets at different stages were extracted to predict the developing themes of public negative opinions regarding racism and xenophobia. The stage-based analysis contributes to the enactment of intervention strategies with an episode-specific focus which will allow for a more effective reduction of the spread of racism and xenophobia on social media under COVID-19.

## Literature Review

In recent years, scholars have extended heated debates around whether the freedom of expression enabled by social media is paving the way for the growth of speech of racism and xenophobia (Jakubowicz 2017; Lara-Cabrera et al. 2017; Thompson 2011). Different from traditional mainstream media, social media empower individuals with the affordance of anonymous speech, which facilitates their engagement in the free expression of opinions beyond institutionalized surveillance and control (Shirky 2011). While prior studies have documented that this affordance could act as an important force to push forward the civil engagement in the pursuit of democracy (Kow, Kou, Semaan, and Cheng 2016; Papacharissi and de Fatima Oliveira 2012; Yamamoto, Kushin, and Dalisay 2015), it may also sustain the rise of anonymous racist propaganda (Ben-David and Matamoros-Fernández 2016; Farkas, Schou, and Neumayer 2018). Furthermore, the promotion of boundless interactions on social media enables the connections between users across geographical spans, consequently allowing for the racist propaganda to reach audiences from far corners of the world and clustering users with a similar mind in large scope and at a high speed (Klein 2017; Jakubowicz 2017).

An increasing number of scholars have switched attention to the flourish of racism and xenophobia on social media (Alkiviadou 2019; Ben-David and Matamoros-Fernández 2016; Chinnasamy and Manaf 2018). Matamoros-Fernández (2017) coined the term "platformed racism" to describe a thriving new form of racism on social media which is constructed and constantly amplified via users' appropriation of affordances enabled by social media platforms. This "platformed racism" often contributes to the reproduction of off-line social inequalities in online public spheres. In her research context, it led to the mediation and circulation of an Australian race-based controversy on Twitter, Facebook, and YouTube. Similarly, Oboler (2016) used the term "Hate 2.0" to refer to the combination of hate speech contents and Facebook as a new way of spreading Islamophobia. Oboler argued that the mode of "Hate 2.0" allows for a large-scale diffusion of biased and hate speeches against the Islam community on social media, which usually makes racist discourse appear as a collective mood. This further helps justify racism as a normal part of public opinions in the online environment, which in return may enable another round of racist thinking diffusion at a large scale on social media.

Our study switches the attention to the growth of racism and xenophobia on social media during COVID-19, which has become one of the most severe public health crises of human beings since the turn of the new millennium. It is well documented that crisis can act as the fuse to trigger the explosion of racism and xenophobia (Coluccello and Kretsos 2015; Salaita 2006). The unpredictable nature of crisis is likely to press the "panic buttons" (Gopalkrishnan 2013) of people who are faced with threats to their health, rights and interests, and even life security. The unstable emotional status has a higher chance to contribute to the subsequent irrational decision-making of the public (Jin and Pang 2010; Sweeny 2008). Especially, the unexpected development of the crisis (Gordon 2003), wide diffusion of misleading online information (Khaldarova and Pantti 2016; Jones 2019), and agenda-setting via media coverage that aims to influence the public opinions for a specific political purpose (Iyengar and Simon 1993), may work together to affect the judgment made by the public during the crisis.

In previous cases, one of the irrational decisions could be the massive blame towards the (im)migrants of racial minorities to negatively view them as the cause of the crisis (Coluccello and Kretsos 2015; Salaita 2006). This bias can arouse a rapid growth of racism and xenophobia amid the public. For instance, Salaita (2006) noted the 9/11 crisis was followed by a rapid transformation of covert racism into overt racism and expressed in actions that impacted adversely on minority Muslim groups in the society. According to Coluccello and Kretsos (2015), social exclusion became the response from the Greek polity and society to irregulated migrants under the stress created by 2008 economic crisis in Greece.

This study examines the racism and xenophobia under the global outbreak of COVID-19. As a disease that is suspected to be originated from China, COVID-19 has been witnessing the rise of discrimination and social exclusion against Chinese, which has been extended to Asians at large. This worldwide phenomenon has been informally defined as "sinophobia". Against this backdrop, social media platforms have become the fields for public expressions of racism and xenophobia (Depoux et al. 2020) which are usually marked by racist hashtags that conflate COVID-19 with a specific racial or national identity. It is within this context that our attention is drawn to the tweets containing racist hashtags, especially, the ones delivering a negative sentiment. Prior studies have noted that race-based discrimination and(or) blame on (im)migrant communities tend to be delivered through the negative sentiment, such as anger, anxiety, and fear, compared to the positive and neutral ones (Redlawsk, Tolbert, and McNeely 2014; Tulkens et al. 2016). Therefore, to understand racism and xenophobia on social media, it becomes necessary to recognize the negative tweets marked with racist hashtags, and the opinions embodied in these tweets. This becomes the focus of this study.

Given the dynamic nature of public sentiments and opinions that tend to change along with the development of crisis (Williamson 1989), this study proposes a stage-based analysis. To specify, this study divides the development of COVID-19 into three major stages – (1) domestic epidemic that was known to the world since the official announcement of Wuhan lockdown by the Chinese government on 23rd Jan, to (2) public health emergency of international concerns since the declaration made by World Health Organization (WHO) on 31st Jan, and later to (3) global pandemic according to the new definition of this disease from WHO on 11th March. With the stage-based approach, we propose the following two research questions:

**RQ1:** Across the three development stages of COVID-19, how does the negative sentiment change in tweets with racist hashtags?

**RQ2:** Across the three development stages of COVID-19, what are the themes of the expression of negative sentiment in the tweets with racist hashtags?

## Methods

### Data Collection and Pre-processing

#Chinesevirus and #Chinavirus were selected as the two most representative racist hashtags for extracting the relevant tweets. Through Twitter API for Python[1], we extracted in total 174,488 tweets in English language marked with either or both of the above two hashtags created between 24th Jan to 31st March in the year of 2020. Repetitive tweets as well as retweets were excluded as they might fail in reflecting a new opinion. Additionally, tweets under 15 words were removed as they are too short for an explicit expression of a viewpoint. After this processing, 69470 tweets constitute the final dataset of this study, with 3,109 tweets falling into the first stage, 6,047 into the second, and 60,314 into the third. It is worth noting a sudden escalation of the numbers of tweets that contain the above hashtags since 16th March and afterwards, which explains the higher number of tweets in the third stage.

### Sentiment Analysis

To answer **RQ1**, we employ VADER for sentiment analysis. With the full name as Valence Aware Dictionary for Sentiment Reasoning, VADER is a popular lexicon-based approach that deploys a rule-based model to categorize the sentiments reflected in texts into, negative, neutral, and positive (Hutto, Clayton, and Gilbert 2014). VADER has been successfully adopted to detect the sentiment in the context of social media (Rodríguez, Argueta, and Chen 2019). Especially, the sentiment categorization by VADER takes consideration of not only the literal meaning of the words, but also negation, punctuation, slangs, and even emoticons, which makes it more suitable for sentiment analysis of the informal content of micro-blogs in comparison with the other popular tools, such as Linguistic Inquiry and Word Count (LIWC), General Inquirer (GI), and SentiWordNet (Bonta, Kumaresh, and Janardhan 2019; Elbagir and Yang 2019).

We use the opensource Python package of VADER[2], operating the model on the 69,470 tweets in the dataset. Each tweet returns four scores as described in the VADER repository -> {'neg', 'pos', 'neu', 'compound'}. The neg, pos, neu scores correspond to the proportion of the text that falls into each category and sum up to 1.0, whereas the compound score is the most important metric which can be used as a unidimensional measure for classifying the sentiment of the tweet. The value of the compound score ranges between -1 (extremely negative) and +1 (extremely positive). We select a commonly used threshold of ±0.05 (Karim & Das, 2018) for the classification of the tweet as negative, positive, or neutral based on the compound score. So, if the compound score of a tweet is greater than or equal to +0.05, it is classified as positive; if it's less than or equal to -0.05 then it is classified as negative; and if it lies between -0.05 and +0.05, it is classified as neutral. Note that before passing a tweet to VADER, we only remove any URLs (http and https) contained in the tweet. We refrain from removing the punctuation marks and emoticons, as VADER makes use of these features for analyzing the sentiment. To check the performance of VADER, we manually classified 1000 randomly selected tweets from our dataset. The overlapping scores between the manual work and VADER reached 76.4%. After scoring and categorizing the sentiment of each tweet, we separated them into the earlier described three stages of COVID-19 according to their date to demonstrate the sentiment changes reflected across the timeline. In summary, in the first stage, the numbers of tweets that classified into negative, neutral and positive sentiments are respectively 1556, 707, 846; in the second stage, 3029, 1243, 1775; in the third stage, 31438, 8569, 20307.

### Keyword Extraction

We make use of keyword extraction to answer **RQ2**. It enables us to capture the thematic words embodied in the ex-pression of negative sentiment in tweets with racist hashtags. Under the extraction, we first removed punctuations, and stop words via making use of Natural Language Toolkit (NLTK)[3] of Python. In a subsequent step, we deleted hashtags and nonsense words, in order to make sure the selected keywords can give us clues about the themes of public discussions with negative sentiment. We eventually extracted the top 50 most frequently used terms for each stage, and word clouds were created accordingly for the visualization of these thematic words.

---

[1] https://github.com/tweepy/tweepy

[2] https://github.com/cjhutto/vaderSentiment

## Findings

### Dominance and Turbulence of Negative Sentiment

In accordance with Fig. 1, negative sentiment keeps on dominating the tone of tweets marked with racist hashtags across the development of COVID-19. In Fig. 1, the proportions of sentiment categories are represented on the y-axis and the three stages of COVID-19 on the x-axis. The ratios of negative sentiment reached 50%, 50%, and 52% respectively at the three stages, staying equal to or greater than the sum of positive and neutral sentiments. In general, we notice that the transformation of COVID-19 into pandemic might act as a watershed, after which the negative expressions concerning racism and xenophobia on Twitter showed a slight rising tendency.

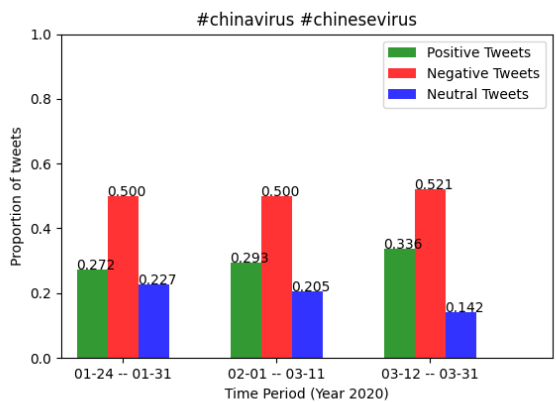

Fig. 1. Proportion of tweets amongst the three categories of sentiment for the three stages (time period) of COVID-19

Fig. 2 illustrates the fluctuation of negative sentiment based on posting dates. In Fig. 2, we plot the proportions of the sentiment categories – positive, negative and neutral on the y-axis for each day of all the three stages on the x-axis. Only the tweets that had posting dates are used and the days that had an extremely small number of tweets were deleted to avoid misleading fluctuating values. From this chart, we can notice the second stage appeared as the most turbulent one where the distance between the peak and the lowest point almost reaches 23%, in comparison with 6% at the first and 9% in the third stage.

At the second stage, the sudden transformation of COVID-19 from a domestic epidemic within the national border of China into an international public health crisis might be the reason that contributes to the turbulence of negative expressions regarding racism and xenophobia on Twitter. To specify, first, the announcement made by WHO that defined COVID-19 as the public health emergency of international concerns alerted the world with the highly contagious nature of the disease, which possibly led to the mood swings of the public. In addition, the emergence of an increasing number of cases in multiple areas globally also played a part in strengthening the anxieties and fears of the public regarding the potential threats imposed by COVID-19 on their health and life security. This is consistent with prior studies that the unpredictable nature of crisis can easily position people into an unstable emotional state (Gopalkrishnan 2012).

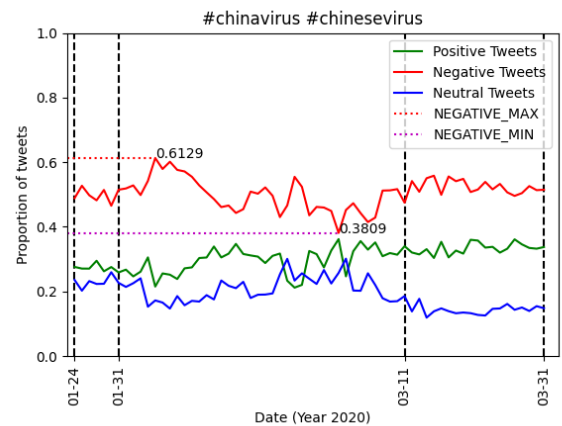

Fig. 2. Distribution of tweets of each sentiment for each day across the three stages of COVID-19

Fig. 2 shows the peak of negative sentiment occurred around the date of 4th Feb. At the peak, the ratio of negative tweets with racist hashtags reaches approximately 61%. This happened a few days after COVID-19 became international public health emergency. In these few days, several countries, such as Singapore, Australia, India, the United States, issued the policy of restricting the entry of tourists who had visited China in the past two weeks. This type of policy was only waived to permanent residents and citizens, and in a few cases, had exemptions made to foreigners who hold a long-term visit pass. The actions of blocking the national borders and excluding foreigners were likely to reinforce xenophobia. Meanwhile, the lockdown policy with a specific country as the target to some degree justified the behaviors of blaming the global spread of COVID-19 on it, which might encourage the growth of racism against people sharing either the racial or national identity with this country. The above-noted factors might result in the rise of negative sentiment when people shared their opinions concerning racism and xenophobia on social media.

---
[3] https://www.nltk.org/

The lowest score of negative sentiment happened around 5th March which is nearly 38%. From the end of February till 5th March, several countries in succession experienced the outbreak of COVID-19, such as South Korea, Italy, Spain, and Iran. The global expansion of COVID-19, on the one hand, distracted the public attention from one specific country to multiple seriously infected nations. This might reduce single country-oriented discrimination and blame, and meanwhile create the necessity of building up a rapport to combat the disease of all human beings. On the other hand, outbreak of COVID-19 in multiple countries led to an increasing body of reports with information regarding this virus from varied sources. The diversity of information might engage people in rethinking about the source and the possible transmission of this novel disease, which tended to trigger their reflections upon their previously biased opinions. These two reasons might cause the declined negative sentiment in tweets when people talked about the association between the coronavirus disease with a particular race or nation. It is worth noting that the declination only happened in the initial phase of global large-scale occurrence of COVID-19, and when the massive diffusion of coronavirus continued, the negative sentiment took off again in tweets with racist hashtags, which can be seen in the third stage of COVID-19.

### Developing Themes of Negative Sentiment

Fig. 3 visualizes the word clouds consisting of terms that were most frequently used for the expression of negative sentiment in tweets marked with racist hashtags at the three development stages of COVID-19. The relative frequency of each term is indicated by its size within the cloud. Table. 1 presents the top 10 terms for each stage. As can be seen, "china" "coronavirus" "world" "chinese" "people" are the common keywords shared by the three stages. It is not difficult to interpret their popularity. COVID-19, a novel coronavirus, concerns the health of people throughout the world. China is the country where the earliest report of the large-scale outbreak of COVID-19 came from, and Chinese is the racial and national identity associated with China. This explains why a number of racist speeches might target at China and Chinese.

Interestingly, apart from the five shared keywords, the rest of keywords demonstrate possibly developing themes across the development of COVID-19. In accordance with Table. 1, at the first stage, public attention was largely drawn to the "outbreak" of COVID-19. And "wuhan", the first city that experienced the outbreak of this disease, was accordingly placed in the spotlight. Meanwhile, negative discussions mainly centered on the (potential) threats arising from COVID-19 to public "health", which might cause "death" and "toll". Therefore, at the first stage when COVID-19 just started its global diffusion, we predict fear might be the key emotional state shared by people who left negative comments with racist hashtags on social media.

At the second stage when multiple countries have experienced the outbreak of COVID-19, we notice new keywords emerged as "infected" and "ccp". The infection of this disease became the thriving theme of negative sentiment, as an increasing number of people sensed the highly contagious nature of COVID-19 along with its global expansion. "ccp", the abbreviation of Communist Party of China, as another rising keyword, to some degree, predicts public discussions around this public health event gradually went political. The arguments of racism and xenophobia on social media might start to involve attacks on political systems.

At the third stage where COVID-19 has been defined as pandemic, there are two major transitions of themes. First, as shown in Table.1, "realdonaldtrump", "us", and "trump" emerged in the list of top 10 keywords. This indicates that a U.S. oriented discourse has been brought to the center of the negative expression of opinions concerning race on social media. This phenomenon might be highly relevant to the speech made by President Trump on 16th March 2020 that openly referred to coronavirus as Chinese virus. This finding is consistent with our data that the number of tweets with hashtags of #Chinavirus and #Chinesevirus rocketed after Trump's speech. Second, it is the first time that "racist" became the keyword. At the same time, the term - "chinaliedpeopledied" which overtly denoted a strong suspicion of and blame on a specific country became common. From this, we predict that there might be an increasing number of users that start to openly deliver negative messages of race-based discrimination and blames when COVID-19 became global pandemic.

### Policy Suggestions

The stage-based approach proposed by our study will enable more effective intervention strategies with a specific focus for each stage under the development of COVID-19. Based on the findings, we provide the following policy suggestions in terms of combatting the growth of racism and xenophobia on social media.

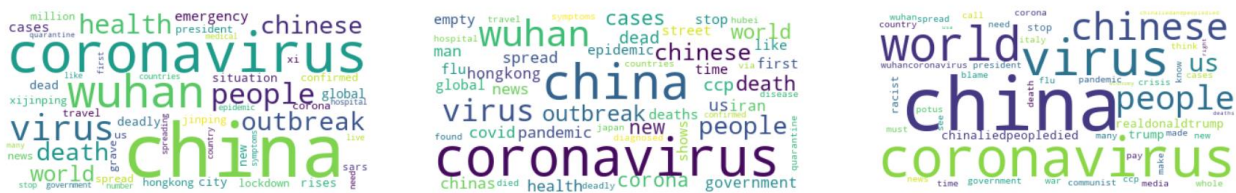

Fig. 3. From left to right – Word clouds of the top 50 most frequently occurring keywords across the three stages of COVID-19

Table 1. Table showing the top 10 most frequently occurring words across the three stages of COVID-19

| Stage 1: 01-24 to 01-31 | | Stage 2: 02-01 to 03-11 | | Stage 3: 03-12 to 03-31 | |
|---|---|---|---|---|---|
| Keyword | Frequency | Keyword | Frequency | Keyword | Frequency |
| china | 1264 | coronavirus | 2367 | china | 14033 |
| coronavirus | 1262 | china | 2039 | coronavirus | 6799 |
| wuhan | 830 | wuhan | 1021 | world | 5599 |
| people | 309 | people | 429 | chinese | 5474 |
| health | 257 | chinese | 375 | people | 5291 |
| outbreak | 234 | outbreak | 346 | us | 2760 |
| death | 211 | death | 266 | chinaliedpeopledied | 2497 |
| chinese | 207 | world | 258 | realdonaldtrump | 2406 |
| toll | 187 | infected | 228 | trump | 2305 |
| world | 173 | ccp | 185 | racist | 2105 |

To specify, in the first stage, when the source and transmission paths of the virus remained to be highly unclear, fears about the virus might be one crucial source where people's race-based bias and blame came from. As noted, whether the virus might cause death was a major concern. Therefore, pacifying the public with the target of reducing their fears can be the focus at the initial stage of public health crisis. One effective strategy will be providing the public with detailed reports regarding the death cases caused by COVID-19 to give them a clearer idea of the lethality of this disease.

At the second stage, first, infectivity of the virus became a new point that might arouse negative reactions from the public, which could be presented in the form of alienation from Chinese and Asians in general. Therefore, it would be helpful to promote the agenda that calls for distinguishing the virus from a specific race or nation instead of conflating them together. Second, as noted, negative sentiment tended to arise from attacks on the political system. Due to this, it became necessary to take action to prevent public discussions around this public health event from going political and maintain the public focus on the nature of the disease.

At the third stage, we can notice the possible negative effects brought about by the racist label declared by the influential public figure on social media. Also, racism and xenophobia could come from suspicions and distrust between countries. Within this context, it is urgent to promote cross-nations transparent information sharing and collaborations to combat fake news.

## Discussion and Conclusions

Situated in the global outbreak of COVID-19, our study enriches the scholarship concerning the emergent racism and xenophobia on social media. We probe into how the freedom of speech on social media, during this public health crisis, might turn into the weapons for the spread of racial exclusion. Our analysis emphasizes the tweets marked with racist hashtags, especially those expressing negative sentiment, as these negative tweets are more likely to deliver the messages containing race relevant bias, discrimination, and blames. Sentiment analysis enables us to capture these negative tweets, and keyword extraction allows for the discovery of themes in their expression of negative sentiment under racist hashtags.

Especially, such analysis is conducted through a stage-based approach. Rather than building the analysis upon the whole set of data, our study proposes to separate the analysis into different stages of COVID-19. In so doing, we are able to map the changing negative sentiment and opinions in tweets with racist hashtags along with the development of COVID-19 from a domestic epidemic to an international public health emergency and later to the global pandemic. It is important to note that the social meanings of this stage-based approach can be well generalized beyond COVID-19. It provides a necessary angle for sentiment analysis in any context of crisis, as the unpredictable nature of crisis often involves people into an unstable emotional state that may lead to their changing opinions regarding their situated crisis (Gopalkrishnan 2013).

Beyond the focus of this study, there are other interesting angles to explore in further research. First, our study limits the analysis to the platform of Twitter. Future research can include other social media platforms in analysis, such as Facebook and Reddit. The expression of racism and xenophobia may vary due to the different technological affordance allowed by each platform. Second, we suggest future research take the social-cultural context into consideration. It will be an innovative angle to compare the expression of racism and xenophobia on social media across countries. Additionally, rather than restricting the analysis to English micro-blogs, future research can extend the analysis of micro-blogs written in other languages. We believe studies across different contexts will contribute to a more comprehensive understanding of the emergent racism and xenophobia on social media platforms in the context of COVID-19, which will help us come up with better strategies to counteract the growth of bias, discrimination, and even hatred under this public health crisis of human beings.